\documentclass[aps,prb,superscriptaddress,reprint,floats,citeautoscript,floatfix,nobibnotes,nofootinbib]{revtex4-1}

\usepackage[intlimits,tbtags]{amsmath}
\usepackage{color}
\usepackage{amssymb}
\usepackage{bm}
\usepackage{array}
\usepackage{dcolumn}
\usepackage{bm}
\usepackage[T1]{fontenc}
\usepackage[]{graphicx}
\DeclareGraphicsExtensions{.pdf}
\usepackage{booktabs}
\usepackage{color}
\usepackage{lineno,hyperref}
\usepackage{soul}
\usepackage{amsmath,bm,amssymb}
\usepackage{subfigure}
\usepackage{setspace}
\usepackage{braket}
\usepackage{mhchem}

\usepackage{textcomp} 
\usepackage{siunitx}  
\DeclareSIUnit\permille{\text{\textperthousand}}

\newcolumntype{d}[1]{D{.}{.}{#1} }

\newcommand{\ilm}{Univ Lyon, Universit\'e Claude Bernard Lyon 1, CNRS,
  Institut Lumi\`ere Mati\`ere, F-69622 Lyon, France}
 \newcommand{\jiangsu}{School of Physics and Electronic Engineering, Jiangsu Normal University, Xuzhou 221116,China}
\newcommand{\jena}{Institut f\"ur Festk\"orpertheorie und -optik,
  Friedrich-Schiller-Universit\"at Jena, Max-Wien-Platz 1, 07743 Jena, Germany}
\newcommand{\halle}{Institut f\"ur Physik, Martin-Luther-Universit\"at
  Halle-Wittenberg, D-06099 Halle, Germany}
\newcommand{\etsf}{European Theoretical Spectroscopy Facility}
\newcommand{\soleil}{Synchrotron SOLEIL, L'Orme des Merisiers, Saint-Aubin, BP48, 91192 Gif-sur-Yvette Cedex, France}
\newcommand{\impmc}{IMPMC-CNRS UMR 7590, Sorbonne Universit\'e, B115, 4 place Jussieu, F-75252 Paris Cedex 05, France}
\newcommand{\ehime}{Geodynamics Research Center, Ehime University, Matsuyama, Ehime 790-8577, Japan}
\newcommand{\tokyo}{Earth-Life Science Institute, Tokyo Institute of Technology, Tokyo 152-8550, Japan}

\begin{document}
\title{Iodine molecule modifications with high pressure}

\author{Jingming Shi}
\affiliation{\ilm}
\affiliation{\jiangsu}
\author{Emiliano Fonda}
\affiliation{\soleil}
\author{Silvana Botti}
\affiliation{\jena}
\affiliation{\etsf}
\author{Miguel A. L. Marques}
\affiliation{\halle}
\affiliation{\etsf}
\author{Toru Shinmei}
\affiliation{\ehime}
\author{Tetsuo~Irifune}
\affiliation{\ehime}
\affiliation{\tokyo}
\author{Anne-Marie Flank}
\affiliation{\soleil}
\author{Pierre Lagarde}
\affiliation{\soleil}
\author{Alain Polian}
\affiliation{\soleil}
\affiliation{\impmc}
\author{Jean-Paul Iti\'e}
\affiliation{\soleil}
\author{Alfonso San-Miguel}\email{alfonso.san-miguel@univ-lyon1.fr}
\affiliation{\ilm}
\date{\today}

\begin{abstract}
Metallization and dissociation are key transformations in diatomic molecules at high densities particularly significant for modeling giant planets. Using X-ray absorption spectroscopy and atomistic modeling, we demonstrate that in halogens, the formation of a \textit{connected} molecular structure takes place at pressures well below metallization. Here we show that the iodine diatomic molecule first elongates of $\sim$0.007 \AA~up to a critical pressure of $P_c$ $\backsim$7~GPa developing bonds between molecules. Then its length continuously decreases with pressure up to 15-20~GPa. Universal trends in halogens are shown and allow to predict for chlorine a pressure of 42$\pm$8~GPa for molecular bond-length reversal. Our findings tackle the molecule invariability paradigm in diatomic molecular phases at high pressures and may be generalized to other abundant diatomic molecules in the universe, including hydrogen.
\end{abstract}
\pacs{}
\maketitle

Diatomic molecules play a fundamental role in life on Earth and have also a prominent presence in other planetary systems. In fact, for solar Jovian planets, as well as for exoplanets~\cite{spiegel-2014}, a key ingredient for the description of their interior is how the structure of diatomic molecules like \ce{H2} or \ce{N2} evolves with pressure.  From a more fundamental point of view, the study of diatomic molecular systems under high pressure contributes to our global understanding of how chemical bonds are modified with increasing atomic density~\cite{johnson2000structure,san2007new,zeng2008new}.
One of the most studied problems concerns the metallization of these molecular solids, which is supposed to occur under sufficiently high pressure. Naturally, the most researched system, and also the most difficult to study, is hydrogen, whose metallization remains the Grail of this field~\cite{klug11}. A very recent infrared study suggests that solid molecular hydrogen becomes metallic at 425~GPa at ambient temperature~\cite{loub20}, while as a fluid it becomes a metal at much lower pressure, namely 140~GPa, under shock compression~\cite{weir96}.



\begin{figure}[htp]
	\centering
	\includegraphics[width=0.95\columnwidth,angle=0]{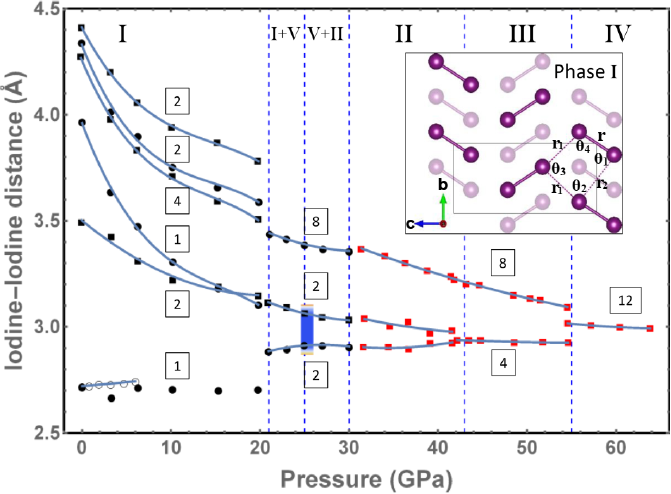}
	\caption{Evolution with pressure of the iodine-iodine distances for the first 12 neighbors through different phases from X-ray diffraction data (filled symbols)\cite{take82,fuji87} and from EXAFS data (hollow symbols)\cite{buon98}. The proposed distance distribution in the modulated inconmensurate phase V, as proposed by Ref.~\onlinecite{kenichi2003modulated}, is shown by a bar at $\sim$25~GPa. Numbers in boxes are the coordination numbers. The inset shows the low-pressure $Cmca$ structure of halogens. The light colored atoms are in the second layer. The symbols $r$, $r_1$ and $r_2$ represent the intramolecular bond-length, first-neighbor and second-neighbor intermolecular distances, respectively, and $\theta_1$, ..., $\theta_4$ the angles shown in the figure.}
		\label{structure}
\end{figure}


The above mentioned molecules (\ce{H2}, \ce{N2} and \ce{O2}) added to the halogens (\ce{F2}, \ce{Cl2}, \ce{Br2} and \ce{I2}) complete the list of the seven homonuclear diatomic molecules stable on the surface of the Earth. Halogens constitute a very interesting class of diatomic molecules, as dissociation and metallization can be reached at much lower pressures for the two heavy ones, \ce{Br2} and \ce{I2}~\cite{take80,fuji89,fuji95}, while they were measured at 200~GPa (metallization) and 260~GPa (dissociation) for chlorine~\cite{dall19}. In addition, halogens appear to follow a common pattern of pressure induced phase transitions providing an excellent benchmark to explore particular features in the evolution of diatomic molecules under increasing density.

In particular, halogens constitute the ideal diatomic molecules to answer questions which have not been satisfactorily addressed to date: How do diatomic molecules evolve before dissociation? Are diatomic molecules invariable during their compression in the molecular phases?

Sofar few studies of molecular systems have addressed this question, and in particular, in halogens, a rigid molecule has been considered in most structural works since earlier studies~\cite{shimomura78}. Nevertheless molecules as C$_{60}$ fullerenes were shown to slightly deform during compression\cite{poloni-2008}. In deuterated water, the D-O bond was shown to increase at a rate of 4~10$^{-4}$~\AA~GPa$^{-1}$ during the compression of ice VIII over the range from 2 to 10~GPa~\cite{nelmes-1993}. Some studies started to tackle the high pressure molecular invariability in halogenes. Iodine molecules evolve to different polyanion arrangements when iodine filled carbon nanotubes are submitted to high pressures\cite{alvarez-2010}. With the use of X-ray absorption spectroscopy (XAS) the question of the evolution of the intramolecular bonding with increasing density in a diatomic molecule was addressed in Br$_2$~\cite{sanmiguel-2000,san2007new}, which showed at 25~GPa a cusp-type evolution of the intramolecular distance with pressure. In solid iodine, similar studies up to 6 GPa~\cite{buon98} showed that the intramolecular distance increases with pressure, but no cusp-type evolution was observed. For liquid iodine, the intramolecular distance was observed to increase faster with pressure and finally a dissociation around 1000 K and 4.5 GPa was detected~\cite{postorino-1999}. In the present work, we combine XAS experiments and density-functional theory calculations to explore the iodine evolution at higher pressures and ambient temperature. We show that the two heavier halogenes present a similar scheme of intramolecular bond evolution with pressure allowing to predict the high pressure behavior of chlorine.

Iodine, bromine and chlorine crystallize at low pressures within the phase I, $Cmca$ structure as shown in the inset of Fig.~\ref{structure}. This is a lamellar structure where each layer is composed of zig-zag chains of diatomic molecules. We denote the intra-chain distance between molecules $r_1$ and inter-chain distance as $r_2$ as shown in inset of Fig.\ref{structure}. In halogen crystals, X-ray diffraction studies do not detect any evolution of the intramolecular distance. The latter is found to remain constant (2.7~\AA~for iodine) up to the emergence of phase V, which is an incommensurate modulated phase~\cite{duan2007ab,kume2005high,kenichi2003modulated,wu2016anomalous}. On the other hand, XAS data\cite{buon98,postorino-1999} show a steady increase of the iodine molecular distance, but not cusp-type evolution as observed in bromine\cite{san2007new}. Dissociation of iodine, which is considered as associated to phase II, is preceded by metallization. In fact, the lamellar structure of iodine leads to an anisotropic change of regime in the electrical conduction observed at 13 and 18~GPa, perpendicularly and parallel to the layers, respectively~\cite{sanmiguel-2000}. High pressure X-ray diffraction~\cite{take82,fuji87} and XAS data~\cite{buon98} for iodine can be gathered to provide the evolution with pressure of the different interatomic distances up to the 12th neighbor as shown in Fig. \ref{structure}.

Bromine follows an analogous scheme of phase transformations as iodine, but taking place at higher pressures. In particular the transition from phase~I to phase~V is observed at 84~GPa and to the monoatomic phase~II at 115~GPa~\cite{kume2005high}. As mentioned, accurate XAS experiments have shown, in the case of bromine, a cusp-type evolution of the Br--Br intramolecular distance. It is experimentally observed that the intramolecular distance first increases with pressure up to 25$\pm5$~GPa at a rate of 2.5~10$^{-3} $ \AA~GPa$^{-1}$ decreasing then at a slightly lower rate until the phase~I instability~\cite{san2007new}. Bromine metallization is proposed at about 60~GPa from electrical conductivity experiments~\cite{shim96}.
\begin{figure}[htp]
	\centering
	\includegraphics[width=0.89\columnwidth,angle=0]{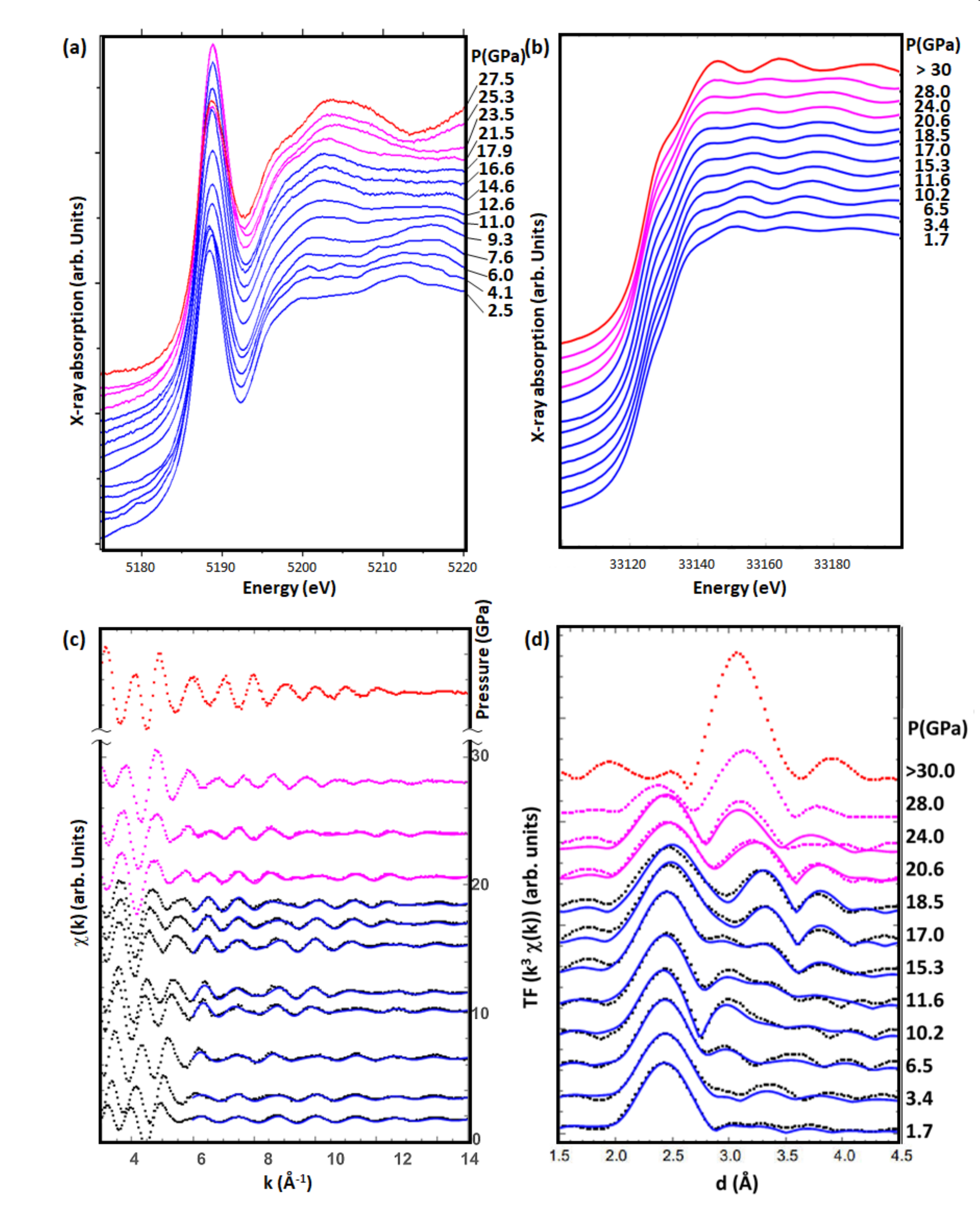}
	\caption{
		\textbf{(a)}~$L_{I}$ XANES spectra of iodine as a function of pressure. \textbf{(b)}~XANES spectra of iodine as function of pressure taken at the K-edge in a different experiment.	\textbf{(c)}~$\chi (k)$ EXAFS signal at the iodine K-edge as function of the photoelectron wavevector $k$ for the same experiment as~(b). The spectra are shifted vertically proportionally to the applied pressure. Pressure corresponds to the baseline and is shown in the right scale of the figure.  \textbf{(d)}~Corresponding pseudo-radial distribution function obtained from the Fourier transform of the EXAFS signal. Blue color corresponds to phase I. Magenta to spectra taken in phase I-V-II mixture region and Red to spectra most likely in region V but with participation of other phases. In (c) and (d) points correspond to the data and lines to the fit (see text for details).}
		\label{allxas}
\end{figure}
 In the following, using also high pressure X-ray absorption experiments, we show that iodine follows a similar cusp-type evolution of its intramolecular distance. We performed 3 different high pressure experiments: a K-edge ($\sim$33.1~keV) XAS study using a diamond anvil cell (DAC) pressure apparatus (we will refer to ``K-DAC'' experiment), a K-edge XAS study using a Paris-Edinburgh press~\cite{mor07} (``K-PE'') and a $L_I$-edge ($\sim$5.2~keV) using a DAC (``L-DAC''). In addition we performed DFT modeling, which pointed to a similar change of behavior as the one observed experimentally. Details of the experiments, data analysis as well as calculation methodologies are provided in the supporting information.



Figures~\ref{allxas}(a) and (b) show the XANES part of the iodine spectra as a function of pressure in experiments done at the iodine $L_{I}$- and K-edges. These two edges allow us to explore the allowed dipolar \textit{s--p} transitions and to put into evidence the so-called ``pre-edge peak'', a prominent peak appearing below the X-ray absorption edge due to the electronic transition from the $1s$ electrons to unfilled 3\textit{p} states associated with the $\sigma$-antibonding molecular orbitals ($\sigma^*$). In the following we will refer to that peak as the $1s \rightarrow 3\sigma^*$ transition. The width of the pre-edge peak increases with pressure up to $\sim$7-10~GPa decreasing then within the $Cmca$ phase (see supplemental material~\cite{suppe}). A change of the pre-edge peak pressure evolution was also observed in bromine at 25$\pm$5~GPa~\cite{sanmiguel-2000}. Because of dipolar selection rules, the structures at the beginning of the iodine K-edge XAS are the projection of the \textit{p}-density of free states modified by the presence of the 1\textit{s} core-hole. In the $L_{1}$ spectra, the pre-edge peak appears well detached, but due to the shorter core-hole lifetime in K-edge, the pre-edge peak appears in that case embedded in the pre-edge region. The $1s \rightarrow 3\sigma^*$ peak characteristics are then related to the iodine electronic structure and it will be worth to compare them to the pressure evolution of the molecular bond evolution, as it has been done for SnI$_4$\cite{Fonda2020}. We will later discuss this aspect. The XANES region shows a progressive evolution with pressure of its different resonances pointing out to a progressive phase transition with phase mixture when evolving from phase I to V at $\sim$20~GPa [Fig. \ref{structure}(b)]. Only the K-edge spectrum labeled ``$>30$~GPa'', as the pressure probe was lost at that pressure, taken most probably in phase II region, shows a very different XANES pattern.

High pressure EXAFS data were acquired in the two different K-edge experiments. Figure~\ref{allxas}(c) shows the EXAFS oscillations as a function of pressure for the K-DAC experiment and panel (d) plots the pseudo-radial distribution function (PRDF) obtained from the Fourier transform of that same data. The lowest distance peak in the PRDF corresponds to the intramolecular distance except for phase-shift corrections. Other interatomic distances are difficult to appreciate in the PRDF at the lowest pressures due to the limited photoelectron mean free path, but they can be progressively observed with increasing pressure. Only the spectrum taken in phase~II region (labeled ``>30 GPa'') shows the disappearance of the initial molecular distance, corroborating the XANES observation of phase mixture between 20 and 30 GPa with presence of phase~I.

The EXAFS analysis could only be properly done on the pure phase,  i.e., below 20~GPa and for the two first spectra in the mixed phase domain for the K-DAC experiment without pressure transmitting medium (PTM). In the K-PE experiment, using BN powder PTM, difficulties on the fit indicated signs of pollution and the data was excluded from the analysis. Details of the EXAFS analysis and enlarged fit figures are given in the supplemental material~\cite{suppe}. Best fits were obtained when fixing all intermolecular structural parameters to thoses provided by X-ray diffraction experiments and leaving the intramolecular distance evolving as only free geometrical parameter. EXAFS fit and their correspondent PRDF are shown as continous lines in Fig.~\ref{allxas}(c) and (d). The obtained experimental distance evolution is shown in Fig.~\ref{distance}(a). More details and the EXAFS data fits are included in the supplemental material (S.M.)~\cite{suppe}.
The obtained evolution with pressure of the intramolecular distance showed to be very robust with respect to details of the fitting procedure (see S.M.). The iodine molecule first expands at a rate of $\sim$\num{1e-3}~\AA/GPa~  up to a pressure of about $\sim$7~GPa. Our data show that at higher pressure the intramolecular distance contracts at a similar rate and returns to its ambient conditions value at $\sim$15-20~GPa. Considering error bars, the maximum amplitude variation of the intramolecular distance in phase-I is then between 0.2 and 0.3\%. The two first data points in the mixed phase could also be correctly fitted using the same method and are shown (color points) in the inset of Fig.~\ref{distance}(a). There is a sudden increase of the fitted intramolecular distance but with a rather low amplitude for what is expected from Fig.~\ref{structure}. We recall that the fit here has been done within a single phase which in case of mixture may give an averaged distance value. The local structure proximity between the two phases and a dominance of phase-I may explain the obtained goodness of fit also for these two points. It is interesting to note here the remarkable correlation between the iodine intramolecular distance evolution with pressure in Fig.~\ref{distance}(a) and the pressure evolution of the FWHM of the $L_{1}$ $1s \rightarrow 3\sigma^*$ electronic transition (see supplemental material~\cite{suppe}). This clearly points out to a strong correlation between the electronic structure evolution of the molecule and its geometry with the antibonding orbitals  progressively hybridazing while I-I elongates. 

\begin{figure}[htp]
	\centering
	\includegraphics[width=1.0\columnwidth,angle=0]{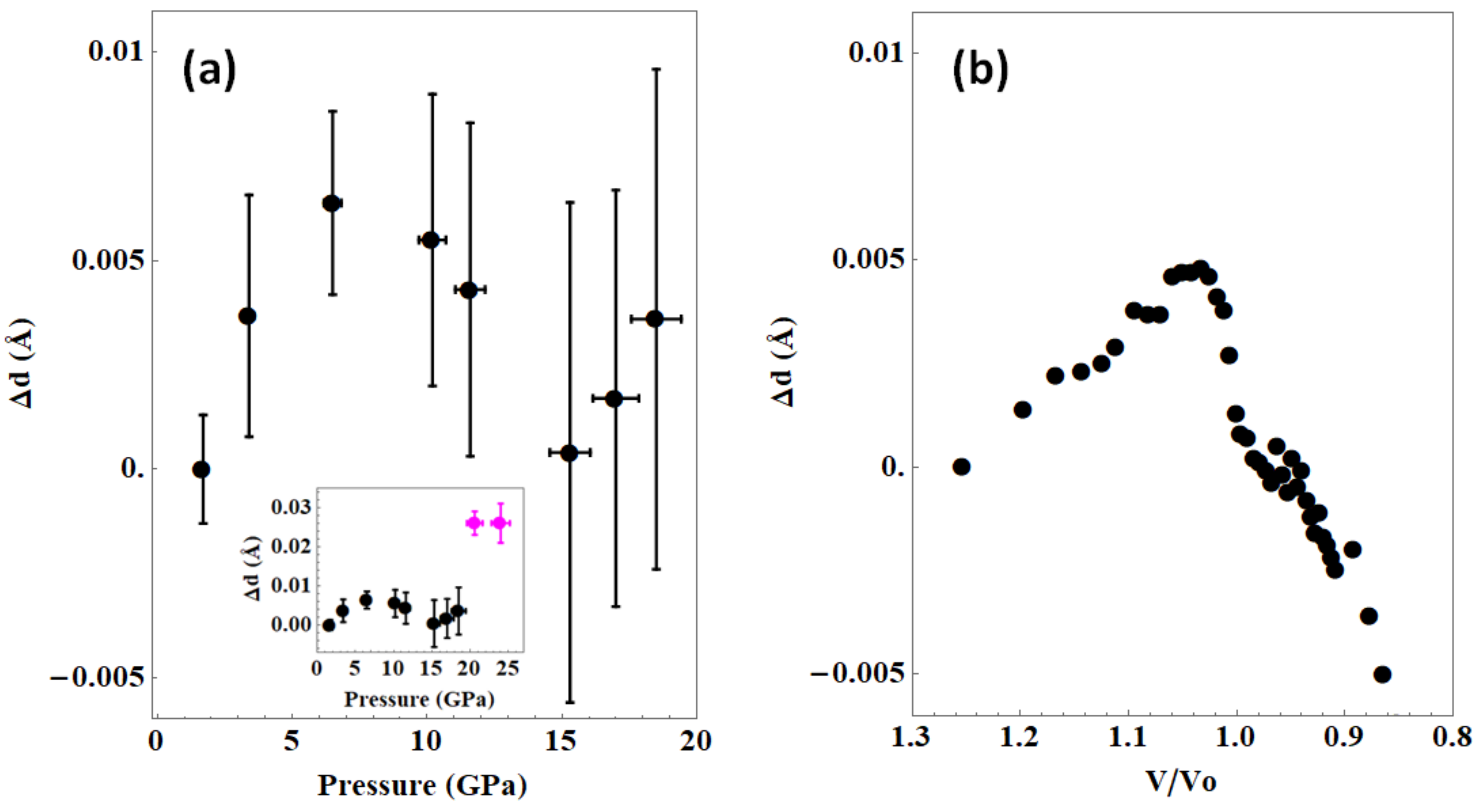}
	\caption{\textbf{(a)}~Evolution with high pressure of the iodine intramolecular distance variation as obtained by EXAFS in the low-pressure $Cmca$ phase. Inset includes also the intramolecular distance for the two first points in the mixed phase. \textbf{(b)}~Variation of the intramolecular distance as a function of the relative volume calculated with DFT. The symbol $V_0$ denotes the equilibrium volume of the crystalline cell at ambient pressure.}
    \label{distance}
\end{figure}

To obtain more insight on the molecular distance evolution with pressure and its connection with the electronic structure changes, we performed DFT simulations. We calculated the lattice parameters and volume of the iodine $Cmca$ structure with different exchange-correlation and van-der-Waals functionals as shown in Tab. S1~\cite{suppe}. By comparing these results with experiments, we chose the most reliable exchange functional for our calculations (see methods section for more details). The lattice parameters that we calculated at ambient pressure ($a = 7.244$~\AA, $b = 4.547$~\AA~ and $c = 9.745$~\AA~, are in very good agreement with the experimental values of $a = 7.103$~\AA, $b = 4.632$~\AA~ and $c = 9.789$~\AA~\cite{ibberson1992high} giving a calculated volume of $321.02$~\AA$^3$ ($322.12$~\AA$^3$ in experiments).  The density of the calculated structure is slightly larger than the experimental one and therefore the theoretical pressures were underestimated. In other words, a smaller calculated ambient volume can be rationalized as the theoretical structure was already pre-compressed. A similar effect was also observed for bromine~\cite{wu2016anomalous}.

\begin{figure*}[t]
	\centering
	\includegraphics[width=0.90\linewidth,angle=0]{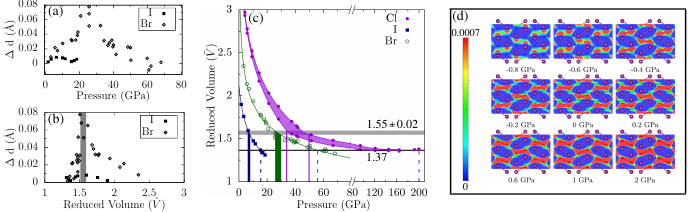}
	\caption{\label{fig:iodine-bromine}\textbf{(a)}~Comparison of intramolecular distance as a function of pressure for iodine (filled symbols) and bromine (empty symbols) as obtained from EXAFS experiments. The bromine data is from Ref.~\onlinecite{san2007new} . \textbf{(b)}~Variation of the intramolecular distance with respect to the corrected scaled volume. The gray bar indicates the location of the critical value for the cusp transition $\tilde{V}_c$. \textbf{(c)} Variation of the corrected scaled volume as function of pressure for iodine, bromine and chlorine. For chlorine The equation of state (EOS) function which presented in Ref.~\onlinecite{dall19} was used to predict the variation of the intramolecular distance (more detailed information can be found in the supplemental material Ref. ~\onlinecite{suppe}). Vertical lines correspond to pressures for the cusp transition, $\tilde{V}_c^{cusp}$=1.55$\pm$0.02 and to the metallization transition, $\tilde{V}_c^{metal}$=1.37. \textbf{(d)}~ Calculated charge density difference plots of $Cmca$ phase of iodine at selected pressures with respected to values of -1 GPa.}
	\label{Br-I}
\end{figure*}
For bromine, EXAFS experiments~\cite{san2007new} revealed that there was a reversal of the evolution of the intramolecular bond length under pressure at 25$\pm5$~GPa. Theoretical simulations observed a similar effect at about 7~GPa~\cite{wu2016anomalous}. In iodine, our experiments show a similar cusp-type behavior at $\sim$7~GPa. To take into account the strong underestimation of this effect by calculations, we performed simulations of iodine in a pressure range from -2~GPa to 20~GPa. Here negative pressures means that the atomic volume is expanded with respect to the volume at atmospheric pressure, $V_0$. The obtained evolution of the atomic volume agrees perfectly with X-ray diffraction data, with all calculated intermolecular distances decreasing with pressure~\cite{suppe}. Figure \ref{distance}(b) shows the intramolecular bond length as a function of the relative volume $V/V_0$ given by the DFT calculations. We observe an anomalous evolution of this quantity which also shows a cusp-type profile and having the same amplitude when compared with experiments. The maximum of the intramolecular bond length is found at around -0.5~GPa which corresponds to $V/V_0\sim$1.03.  

In Fig.~\ref{Br-I}.a we compare our experimental result for the evolution of the iodine intramolecular distance with pressure with the one for bromine~\cite{san2007new}. We observe that both iodine and bromine follow a similar pattern. First the intramolecular distance increases up to a critical pressure and then it decreases. The pressure slopes for the two halogens are difficult to compare due to the limited pressure domain for iodine and to the different regimes in bromine, but overall the pressure rate of variation seems to be more important for bromine. This initially positive slope changes sign at $P_c\sim$7~GPa for iodine and at 25~GPa for bromine. 
We show in Fig.\ref{Br-I}(b) the variation of the intramolecular distance for bromine and iodine in the $Cmca$ phase with respect to the reduced volume $\tilde{V}$. The reduced volume is defined as $\tilde{V}=V/(r^3)$, following Ref.~\onlinecite{fuji89} where $r$ is the intramolecular length. This parameter was shown to be relevant to scale the phase diagrams of the different halogens. We have corrected the reduced volume to take into account the measured pressure dependence of iodine and bromine intramolecular bond-lengths. As shown in Fig.\ref{Br-I}(b), when plotting the intramolecular distance variation in iodine and bromine against the corrected scaled volume, we observe that both curves peak at the same value, which we estimate graphically at $\tilde{V}_c^{cusp}=1.55\pm0.02$. We use this value to make an estimation of $P_c$ for chlorine. In Fig.~\ref{Br-I}c, the corrected reduced volume evolution with pressure is represented for chlorine, bromine and iodine. For chlorine, $\tilde{V}$ is calculated from the atomic volume of Ref.~\onlinecite{dall19}. We estimated chlorine $\tilde{V}$ by finding out the change of the molecular bond-length which allows for a common scaled critical volume at metallization,$\tilde{V}_c^{metal}$, for the three halogens matching experimental values\cite{shim96,sanmiguel-2000,dall19} (see S.I. for details). This leads on one side to an estimate of the corresponding intramolecular cusp pressures in I$_2$, Br$_2$ and Cl$_2$ of respectively 7.1$\pm$0.7~GPa, 27.8$\pm$2.1~GPa and 41.5$\pm$7.5~GPa for $\tilde{V}_c^{cusp}=1.55\pm0.02$ and to respective metallization pressures of 14~GPa, 57~GPa and 200~GPa (this last fixed from Ref. \onlinecite{dall19}) corresponding to $\tilde{V}_c^{metal}=1.37\pm0.02$. We have considered for Chlorine a pressure rate of change of the intramolecular bond up to the cusp pressure comprised between of 2.0~10$^{-3}$ and 4.0~10$^{-3}$~\AA/GPa which represents the shaded region in Fig.~\ref{Br-I}c. 
We note that measured (iodine, bromine) and predicted values (chlorine) of cusp critical pressure correspond to the crossing pressures of the Raman scattering A$_g$(S) and B$_{3g}$(S) zone-center phonon modes~\cite{kume2005high,dall19} which provides robustness to our observations and our prediction for chlorine.
To better understand the physical significance of the observed evolution of the intramolecular distance, in Fig.~\ref{Br-I}.(d) we plot the calculated differential electronic charge density (ECD) in the (100) plane of the $Cmca$ structure, taking as reference the corresponding ECD at -1~GPa. We observe that with increasing pressure the electronic density in the region between the parallel molecules, which corresponds to $r_2$ in Fig.~\ref{structure}, becomes progressively more important. This change is particularly important for the first calculated pressures and is concomitant to weak changes in the valence band electronic density of states (see S.I.), supporting the idea that the pressure-induced dilatation of the I$_2$ molecule may be attributed to a redistribution of the ECD to create intermolecular bonds between parallel molecules and then a \textit{connected molecular structure}. Similar conclusions were proposed for Br$_2$~\cite{wu2016anomalous}. The observed pressure evolution of the $L_{1}$ $1s \rightarrow 3\sigma^*$ electronic transition in correlation with the intramolecular bond modifications also gives support to this explanation.

In conclusion we have shown that the length of the iodine molecule is modified with pressure. With increasing density, the iodine molecule elongates up to a critical pressure of $\sim$~7 GPa within the $Cmca$ phase. From this pressure, the molecule length starts to decrease. The evolution of the scaled volume of I$_2$ and Br$_2$ allows to predict a similar bond-length reversal in the $Cmca$ Cl$_2$ crystal at 42$\pm$8~GPa. Our atomistic modeling shows that the observed behavior may be attributed to an electron charge redistribution to stabilize in-plane intermolecular bonds leading to a connected molecular structure. This bond scheme appears at pressures well below metallization. The possibility that this observed behavior could be extended to other diatomic molecules including hydrogen, constitutes an open question.


\acknowledgements
We would like to thank Dr. Olivier Mathon (ESRF, Grenoble) and Dr. Vittoria Pischedda (University of Lyon) for their assistance in some of the experiments. J.S. acknowledges the Project Funded by the National Natural Science Foundation of China under Grant No. 11804129 and financial support from the China Scholarship Council. Computational resources were provide by x2016096017 Project of Curie. A.S.M acknowledges support from the PLECE platform of the University of Lyon. 


\begin{thebibliography}{10}
	\expandafter\ifx\csname url\endcsname\relax
	\def\url#1{\texttt{#1}}\fi
	\expandafter\ifx\csname urlprefix\endcsname\relax\def\urlprefix{URL }\fi
	\providecommand{\bibinfo}[2]{#2}
	\providecommand{\eprint}[2][]{\url{#2}}
	
	\bibitem{spiegel-2014}
	\bibinfo{author}{Spiegel, D.~S.}, \bibinfo{author}{Fortney, J.~J.} \&
	\bibinfo{author}{Sotin, C.}
	\newblock \bibinfo{title}{Structure of exoplanets}.
	\newblock \emph{\bibinfo{journal}{Proc. Natl. Acad. Sci. U. S. A.}}
	\textbf{\bibinfo{volume}{111}}, \bibinfo{pages}{12622--12627}
	(\bibinfo{year}{2014}).
	
	\bibitem{johnson2000structure}
	\bibinfo{author}{Johnson, K.~A.} \& \bibinfo{author}{Ashcroft, N.}
	\newblock \bibinfo{title}{Structure and bandgap closure in dense hydrogen}.
	\newblock \emph{\bibinfo{journal}{Nature}} \textbf{\bibinfo{volume}{403}},
	\bibinfo{pages}{632--635} (\bibinfo{year}{2000}).
	
	\bibitem{san2007new}
	\bibinfo{author}{San-Miguel, A.} \emph{et~al.}
	\newblock \bibinfo{title}{New phase transition of solid bromine under high
		pressure}.
	\newblock \emph{\bibinfo{journal}{Phys. Rev. Lett.}}
	\textbf{\bibinfo{volume}{99}}, \bibinfo{pages}{015501}
	(\bibinfo{year}{2007}).
	
	\bibitem{zeng2008new}
	\bibinfo{author}{Zeng, Q.} \emph{et~al.}
	\newblock \bibinfo{title}{A new phase of solid iodine with different molecular
		covalent bonds}.
	\newblock \emph{\bibinfo{journal}{Proc. Natl. Acad. Sci. U. S. A.}}
	\textbf{\bibinfo{volume}{105}}, \bibinfo{pages}{4999--5001}
	(\bibinfo{year}{2008}).
	
	\bibitem{klug11}
	\bibinfo{author}{Klug, D.~D.} \& \bibinfo{author}{Yao, Y.}
	\newblock \bibinfo{title}{Metallization of solid hydrogen: the challenge and
		possible solutions}.
	\newblock \emph{\bibinfo{journal}{Phys. Chem. Chem. Phys.}}
	\textbf{\bibinfo{volume}{13}}, \bibinfo{pages}{16999--17006}
	(\bibinfo{year}{2011}).
	
	\bibitem{loub20}
	\bibinfo{author}{Loubeyre, P.}, \bibinfo{author}{Occelli, F.} \&
	\bibinfo{author}{Dumas, P.}
	\newblock \bibinfo{title}{Synchrotron infrared spectroscopic evidence of the
		probable transition to metal hydrogen}.
	\newblock \emph{\bibinfo{journal}{Nature}} \textbf{\bibinfo{volume}{577}},
	\bibinfo{pages}{631--635} (\bibinfo{year}{2020}).
	
	\bibitem{weir96}
	\bibinfo{author}{Weir, S.~T.}, \bibinfo{author}{Mitchell, A.~C.} \&
	\bibinfo{author}{Nellis, W.~J.}
	\newblock \bibinfo{title}{Metallization of fluid molecular hydrogen at 140 gpa
		(1.4 mbar)}.
	\newblock \emph{\bibinfo{journal}{Phys. Rev. Lett.}}
	\textbf{\bibinfo{volume}{76}}, \bibinfo{pages}{1860--1863}
	(\bibinfo{year}{1996}).
	
	\bibitem{take82}
	\bibinfo{author}{Takemura, K.}, \bibinfo{author}{Minomura, S.},
	\bibinfo{author}{Shimomura, O.}, \bibinfo{author}{Fujii, Y.} \&
	\bibinfo{author}{Axe, J.~D.}
	\newblock \emph{\bibinfo{journal}{Phys. Rev. B}} \textbf{\bibinfo{volume}{26}},
	\bibinfo{pages}{998} (\bibinfo{year}{1982}).
	
	\bibitem{fuji87}
	\bibinfo{author}{Fujii, Y.} \emph{et~al.}
	\newblock \bibinfo{title}{Pressure-induced face-centered-cubic phase of
		monatomic metallic iodine}.
	\newblock \emph{\bibinfo{journal}{Phys. Rev. Lett.}}
	\textbf{\bibinfo{volume}{58}}, \bibinfo{pages}{796} (\bibinfo{year}{1987}).
	
	\bibitem{buon98}
	\bibinfo{author}{Buontempo, U.} \emph{et~al.}
	\newblock \bibinfo{title}{Anomalous bond length expansion in liquid iodine at
		high pressure}.
	\newblock \emph{\bibinfo{journal}{Phys. Rev. Lett.}}
	\textbf{\bibinfo{volume}{80}}, \bibinfo{pages}{1912--1915}
	(\bibinfo{year}{1998}).
	
	\bibitem{kenichi2003modulated}
	\bibinfo{author}{Takemura, K.}, \bibinfo{author}{Sato, K.},
	\bibinfo{author}{Fujihisa, H.} \& \bibinfo{author}{Onoda, M.}
	\newblock \bibinfo{title}{Modulated structure of solid iodine during its
		molecular dissociation under high pressure}.
	\newblock \emph{\bibinfo{journal}{Nature}} \textbf{\bibinfo{volume}{423}},
	\bibinfo{pages}{971--974} (\bibinfo{year}{2003}).
	
	\bibitem{take80}
	\bibinfo{author}{Takemura, K.}, \bibinfo{author}{Minomura, S.},
	\bibinfo{author}{Shimomura, O.} \& \bibinfo{author}{Fujii, Y.}
	\newblock \emph{\bibinfo{journal}{Phys. Rev. Lett.}}
	\textbf{\bibinfo{volume}{45}}, \bibinfo{pages}{1881} (\bibinfo{year}{1980}).
	
	\bibitem{fuji89}
	\bibinfo{author}{Fujii, Y.} \emph{et~al.}
	\newblock \emph{\bibinfo{journal}{Phys. Rev. Lett}}
	\textbf{\bibinfo{volume}{63}}, \bibinfo{pages}{536} (\bibinfo{year}{1989}).
	
	\bibitem{fuji95}
	\bibinfo{author}{Fujihisa, H.}, \bibinfo{author}{Fujii, Y.},
	\bibinfo{author}{Takemura, K.} \& \bibinfo{author}{Shimomura, O.}
	\newblock \emph{\bibinfo{journal}{J. Phys. Chem. Solids}}
	\textbf{\bibinfo{volume}{56}}, \bibinfo{pages}{1439} (\bibinfo{year}{1995}).
	
	\bibitem{dall19}
	\bibinfo{author}{Dalladay-Simpson, P.} \emph{et~al.}
	\newblock \bibinfo{title}{Band gap closure, incommensurability and molecular
		dissociation of dense chlorine}.
	\newblock \emph{\bibinfo{journal}{Nat. Commun.}} \textbf{\bibinfo{volume}{10}},
	\bibinfo{pages}{1134} (\bibinfo{year}{2019}).
	
	\bibitem{shimomura78}
	\bibinfo{author}{Shimomura, C.} \emph{et~al.}
	\newblock \bibinfo{title}{Structure analysis of high-pressure metallic state of
		iodine}.
	\newblock \emph{\bibinfo{journal}{Phys. Rev. B}} \textbf{\bibinfo{volume}{18}},
	\bibinfo{pages}{715--719} (\bibinfo{year}{1978}).
	
	\bibitem{poloni-2008}
	\bibinfo{author}{Poloni, R.} \emph{et~al.}
	\newblock \bibinfo{title}{Pressure-induced deformation of the {C}$_{60}$
		fullerene in {Rb}$_{6}${C}$_{60}$ and {Cs}$_{6}${C}$_{60}$}.
	\newblock \emph{\bibinfo{journal}{Phys. Rev. B}} \textbf{\bibinfo{volume}{77}},
	\bibinfo{pages}{035429} (\bibinfo{year}{2008}).
	
	\bibitem{nelmes-1993}
	\bibinfo{author}{Nelmes, R.~J.} \emph{et~al.}
	\newblock \bibinfo{title}{Neutron diffraction study of the structure of
		deuterated ice {VIII} to 10 {GPa}}.
	\newblock \emph{\bibinfo{journal}{Phys. Rev. Lett.}}
	\textbf{\bibinfo{volume}{71}}, \bibinfo{pages}{1192--1195}
	(\bibinfo{year}{1993}).
	
	\bibitem{alvarez-2010}
	\bibinfo{author}{Alvarez, L.} \emph{et~al.}
	\newblock \bibinfo{title}{High-pressure behavior of polyiodides confined into
		single-walled carbon nanotubes: A raman study}.
	\newblock \emph{\bibinfo{journal}{Phys. Rev. B}} \textbf{\bibinfo{volume}{82}},
	\bibinfo{pages}{205403} (\bibinfo{year}{2010}).
	
	\bibitem{sanmiguel-2000}
	\bibinfo{author}{San~Miguel, A.} \emph{et~al.}
	\newblock \bibinfo{title}{Bromine metallization studied by {X}-ray absorption
		spectroscopy}.
	\newblock \emph{\bibinfo{journal}{Eur. Phys. J. B}}
	\textbf{\bibinfo{volume}{17}}, \bibinfo{pages}{227--233}
	(\bibinfo{year}{2000}).
	
	\bibitem{postorino-1999}
	\bibinfo{author}{Postorino, P.}, \bibinfo{author}{Buontempo, U.},
	\bibinfo{author}{Filipponi, A.} \& \bibinfo{author}{Nardone, M.}
	\newblock \bibinfo{title}{Early metallization in molecular fluids: the case of
		iodine}.
	\newblock \emph{\bibinfo{journal}{Phys. B}} \textbf{\bibinfo{volume}{265}},
	\bibinfo{pages}{72 -- 78} (\bibinfo{year}{1999}).
	
	\bibitem{duan2007ab}
	\bibinfo{author}{Duan, D.} \emph{et~al.}
	\newblock \bibinfo{title}{Ab initio studies of solid bromine under high
		pressure}.
	\newblock \emph{\bibinfo{journal}{Phys. Rev. B}} \textbf{\bibinfo{volume}{76}},
	\bibinfo{pages}{104113} (\bibinfo{year}{2007}).
	
	\bibitem{kume2005high}
	\bibinfo{author}{Kume, T.}, \bibinfo{author}{Hiraoka, T.},
	\bibinfo{author}{Ohya, Y.}, \bibinfo{author}{Sasaki, S.} \&
	\bibinfo{author}{Shimizu, H.}
	\newblock \bibinfo{title}{High pressure raman study of bromine and iodine: soft
		phonon in the incommensurate phase}.
	\newblock \emph{\bibinfo{journal}{Phys. Rev. Lett.}}
	\textbf{\bibinfo{volume}{94}}, \bibinfo{pages}{065506}
	(\bibinfo{year}{2005}).
	
	\bibitem{wu2016anomalous}
	\bibinfo{author}{Wu, M.}, \bibinfo{author}{Tse, J.~S.} \& \bibinfo{author}{Pan,
		Y.}
	\newblock \bibinfo{title}{Anomalous bond length behavior and a new solid phase
		of bromine under pressure}.
	\newblock \emph{\bibinfo{journal}{Sci. Rep.}} \textbf{\bibinfo{volume}{6}},
	\bibinfo{pages}{25649} (\bibinfo{year}{2016}).
	
	\bibitem{shim96}
	\bibinfo{author}{Shimizu, K.}, \bibinfo{author}{Amaya, K.} \&
	\bibinfo{author}{Endo, S.}
	\newblock \bibinfo{title}{{High Pressure Science and Technology: Proceedings of
			the Joint XV AIRAPT and XXXIII EHPRG International Conference, Warsaw,
			Poland, 1995}}  (\bibinfo{year}{1996}).
	
	\bibitem{mor07}
	\bibinfo{author}{Morard, G.} \emph{et~al.}
	\newblock \bibinfo{title}{Optimization of parisâ€“edinburgh press cell
		assemblies for in situ monochromatic x-ray diffraction and x-ray absorption}.
	\newblock \emph{\bibinfo{journal}{High Pressure Res.}}
	\textbf{\bibinfo{volume}{27}}, \bibinfo{pages}{223--233}
	(\bibinfo{year}{2007}).
	
	\bibitem{suppe}
	\bibinfo{note}{See Supplemental Material at [URL will be inserted by
		publisher], which includes
		Refs.~[\citenum{Ishimatsu2012Glitch-free,san-miguel-1995,Ankudinov-1998,Newville-2001,giannozzi2009quantum,pbe,klimevs2011van,klimevs2009chemical}].
		It contains more detailed experiment methods, EXAFS analysis, complementary
		figures and computational methods, etc.}
	
	\bibitem{Fonda2020}
	\bibinfo{author}{Fonda, E.}, \bibinfo{author}{Polian, A.},
	\bibinfo{author}{Shinmei, T.}, \bibinfo{author}{Irifune, T.} \&
	\bibinfo{author}{ItiÃ©, J.-P.}
	\newblock \bibinfo{title}{Mechanism of pressure induced amorphization of sni4:
		A combined x-ray diffractionâ€”x-ray absorption spectroscopy study}.
	\newblock \emph{\bibinfo{journal}{The Journal of Chemical Physics}}
	\textbf{\bibinfo{volume}{153}}, \bibinfo{pages}{064501}
	(\bibinfo{year}{2020}).
	
	\bibitem{ibberson1992high}
	\bibinfo{author}{Ibberson, R.}, \bibinfo{author}{Moze, O.} \&
	\bibinfo{author}{Petrillo, C.}
	\newblock \bibinfo{title}{High resolution neutron powder diffraction studies of
		the low temperature crystal structure of molecular iodine (i2)}.
	\newblock \emph{\bibinfo{journal}{Mol. Phys.}} \textbf{\bibinfo{volume}{76}},
	\bibinfo{pages}{395--403} (\bibinfo{year}{1992}).
	
	\bibitem{Ishimatsu2012Glitch-free}
	\bibinfo{author}{Ishimatsu, N.} \emph{et~al.}
	\newblock \bibinfo{title}{{Glitch-free X-ray absorption spectrum under high
			pressure obtained using nano-polycrystalline diamond anvils}}.
	\newblock \emph{\bibinfo{journal}{J. Synchrotron Radiat.}}
	\textbf{\bibinfo{volume}{19}}, \bibinfo{pages}{768--772}
	(\bibinfo{year}{2012}).
	
	\bibitem{san-miguel-1995}
	\bibinfo{author}{San-Miguel, A.}
	\newblock \bibinfo{title}{A program for fast classic or dispersive \{XAS\} data
		analysis in a \{PC\}}.
	\newblock \emph{\bibinfo{journal}{Phys. B}}
	\textbf{\bibinfo{volume}{208-209}}, \bibinfo{pages}{177 -- 179}
	(\bibinfo{year}{1995}).
	
	\bibitem{Ankudinov-1998}
	\bibinfo{author}{Ankudinov, A.~L.}, \bibinfo{author}{Ravel, B.},
	\bibinfo{author}{Rehr, J.~J.} \& \bibinfo{author}{Conradson, S.~D.}
	\newblock \bibinfo{title}{Real-space multiple-scattering calculation and
		interpretation of x-ray-absorption near-edge structure}.
	\newblock \emph{\bibinfo{journal}{Phys. Rev. B}} \textbf{\bibinfo{volume}{58}},
	\bibinfo{pages}{7565--7576} (\bibinfo{year}{1998}).
	
	\bibitem{Newville-2001}
	\bibinfo{author}{Newville, M.}
	\newblock \bibinfo{title}{{{\it IFEFFIT}: interactive XAFS analysis and {\it
				FEFF} fitting}}.
	\newblock \emph{\bibinfo{journal}{J. Synchrotron Radiat.}}
	\textbf{\bibinfo{volume}{8}}, \bibinfo{pages}{322--324}
	(\bibinfo{year}{2001}).
	
	\bibitem{giannozzi2009quantum}
	\bibinfo{author}{Giannozzi, P.} \emph{et~al.}
	\newblock \bibinfo{title}{Quantum espresso: a modular and open-source software
		project for quantum simulations of materials}.
	\newblock \emph{\bibinfo{journal}{J. Phys.: Condens. Matter}}
	\textbf{\bibinfo{volume}{21}}, \bibinfo{pages}{395502}
	(\bibinfo{year}{2009}).
	
	\bibitem{pbe}
	\bibinfo{author}{Perdew, J.~P.}, \bibinfo{author}{Burke, K.} \&
	\bibinfo{author}{Ernzerhof, M.}
	\newblock \bibinfo{title}{Generalized gradient approximation made simple}.
	\newblock \emph{\bibinfo{journal}{Phys. Rev. Lett.}}
	\textbf{\bibinfo{volume}{77}}, \bibinfo{pages}{3865} (\bibinfo{year}{1996}).
	
	\bibitem{klimevs2011van}
	\bibinfo{author}{Klime{\v{s}}, J.}, \bibinfo{author}{Bowler, D.~R.} \&
	\bibinfo{author}{Michaelides, A.}
	\newblock \bibinfo{title}{Van der waals density functionals applied to solids}.
	\newblock \emph{\bibinfo{journal}{Phys. Rev. B}} \textbf{\bibinfo{volume}{83}},
	\bibinfo{pages}{195131} (\bibinfo{year}{2011}).
	
	\bibitem{klimevs2009chemical}
	\bibinfo{author}{Klime{\v{s}}, J.}, \bibinfo{author}{Bowler, D.~R.} \&
	\bibinfo{author}{Michaelides, A.}
	\newblock \bibinfo{title}{Chemical accuracy for the van der waals density
		functional}.
	\newblock \emph{\bibinfo{journal}{J. Phys.: Condens. Matter}}
	\textbf{\bibinfo{volume}{22}}, \bibinfo{pages}{022201}
	(\bibinfo{year}{2009}).
	
\end{thebibliography}
\end{document}